\documentclass[eprint,review=false
]{actapoly}

\usepackage[utf8]{inputenc}

\begin{document}

\title[Correlation between X-ray and gamma data of Swift measurements]
{Correlation between X-ray and gamma data of Swift measurements}

\correspondingauthor[I. I. Racz]{Istvan I. Racz}{my}{racz.istvan@uni-nke.hu}
\author[L. G. Balazs]{Lajos G. Balazs}{third,their}%{third}
\author[I. Horvath]{Istvan Horvath}{my}
\author[S. Pinter]{Sandor Pinter}{my,their}

\institution{my}{University of Public Service, Department of Natural Sciences, 2 Ludovika tér, H-1083 -- Budapest, Hungary}
\institution{third}{HUN-REN Research Centre for Astronomy and Earth Sciences, Konkoly Thege Miklós Astronomical Institute, Konkoly-Thege Miklós út 15-17., H-1121, Budapest, Hungary}
\institution{their}{Eötvös Loránd University, Faculty of Science, Institute of Physics and Astronomy, Department of Astronomy, Pázmány Péter sétány 1/A., H-1117, Budapest, Hungary}

\begin{abstract}
Several studies over the last two decades have used canonical correlation analysis (CCA) to study the relationships between main $\gamma$-ray (e.g. fluence, peak flux, and duration) and main X-ray (flux, decay and spectral index, and hydrogen column density) data from gamma-ray bursts (GRBs). In this paper, we revisit this approach using a much larger dataset to identify potential new insights into these relationships. We used CCA to investigate the interrelationship of the aforementioned gamma-ray and X-ray parameters. Using the derived canonical variables, we calculated their correlations (canonical loadings) with the original data. Consistently with previous research, the analysis revealed that gamma-ray fluence and X-ray flux have the strongest correlation, while the X-ray decay index and spectral index have a lower contribution. Interestingly, our analysis of a much larger dataset reveals that the HI column density makes a significant contribution to the overall correlation. This finding, in the context of the collapsar model for long GRBs, could be interpreted as an indication that the progenitor star ejected an HI envelope during the GRB.

\end{abstract}


\maketitle

\section{Introduction}

Since the Big Bang, the gamma-ray bursts (GRBs) have been the most intense and brightest events in the Universe\cite{meszaros_2006,Kumar}. GRBs, are usually short, powerful bursts of radiation that can last anywhere from a few milliseconds to several minutes. They were first identified in the late 1960s \cite{1973ApJ...185L..57W}. These bursts are important probes of the early Universe because they come from far-off galaxies, frequently billions of light years away.

The main characteristic of gamma-ray bursts is their enormous energy output; some bursts can release more energy in a matter of seconds than the Sun will in a billion years. According to how long they last, GRBs are typically divided into two categories: short GRBs, which last less than two seconds, and long GRBs, which can last anywhere from two to several minutes. Short GRBs are thought to result from neutron star mergers \cite{ns_merging}, while long GRBs are thought to be the result of massive stars collapsing into black holes (the collapsar model) \cite{collapsar, meszaros_fireball}. Evidence for the first case can be found in the connection with gravitational waves, which have already been observed several times \cite{2017PhRvL.119n1101A, veres_gw170814,2016A&A...593L..10B,2017CoSka..47...76B,2018Ap&SS.363...53H}. 

Apart from the two primary models, it was observed in the 1990s and early 2000s that there are other varieties of GRBs \cite{hoi1998, 1998ApJ...508..314M}, which have medium lengths of a few seconds and spectral hardnesses that differ from the two mentioned above \citep{hor04,hoi2006,2020MNRAS.494.4343K}. Although the physical composition of this intermediate group is still unknown, it appears that the intermediate GRBs and X-ray flash events are related \citep{hoi2010, 2010ApJ...725.1955V, pinter_2018, Xiongwei2018}. Apart from the initial gamma-ray emission, Gamma-Ray Bursts (GRBs) often emit an afterglow that radiates at different wavelengths, such as X-rays, ultraviolet, optical, and radio waves. This afterglow can yield important insights into the surroundings and mechanisms of these explosive events \cite{zhang_2018, 2019MNRAS.486.4823T}.

\section{Materials and methods}

\subsection{Data}

The Neil Gehrels Swift Observatory (formerly Swift telescope) is a multi-wavelength observatory that focuses on studying gamma-ray burst (GRB) science. It uses three instruments to observe GRBs and afterglows in gamma-ray, X-ray, ultraviolet, and optical wavebands. The main objectives are to determine the origin, classify bursts, and study their evolution and interaction with the environment. Swift discovers around 100 bursts per year and provides accurate position estimates, multi-wavelength lightcurves, gamma-ray spectrum, and X-ray spectra. Launched in 2004 as part of NASA's MIDEX program, Swift is the most comprehensive study of GRB afterglows to date.

The Burst Alert Telescope (BAT) is designed to provide key GRB triggers and a draft location with an error of about 4-arcmin. The imaging energy range is 15-150 keV with a non-coded response of up to 500 keV. 

\begin{figure}
\centering
\includegraphics[width=1.\linewidth, trim= -25 0 100 0]{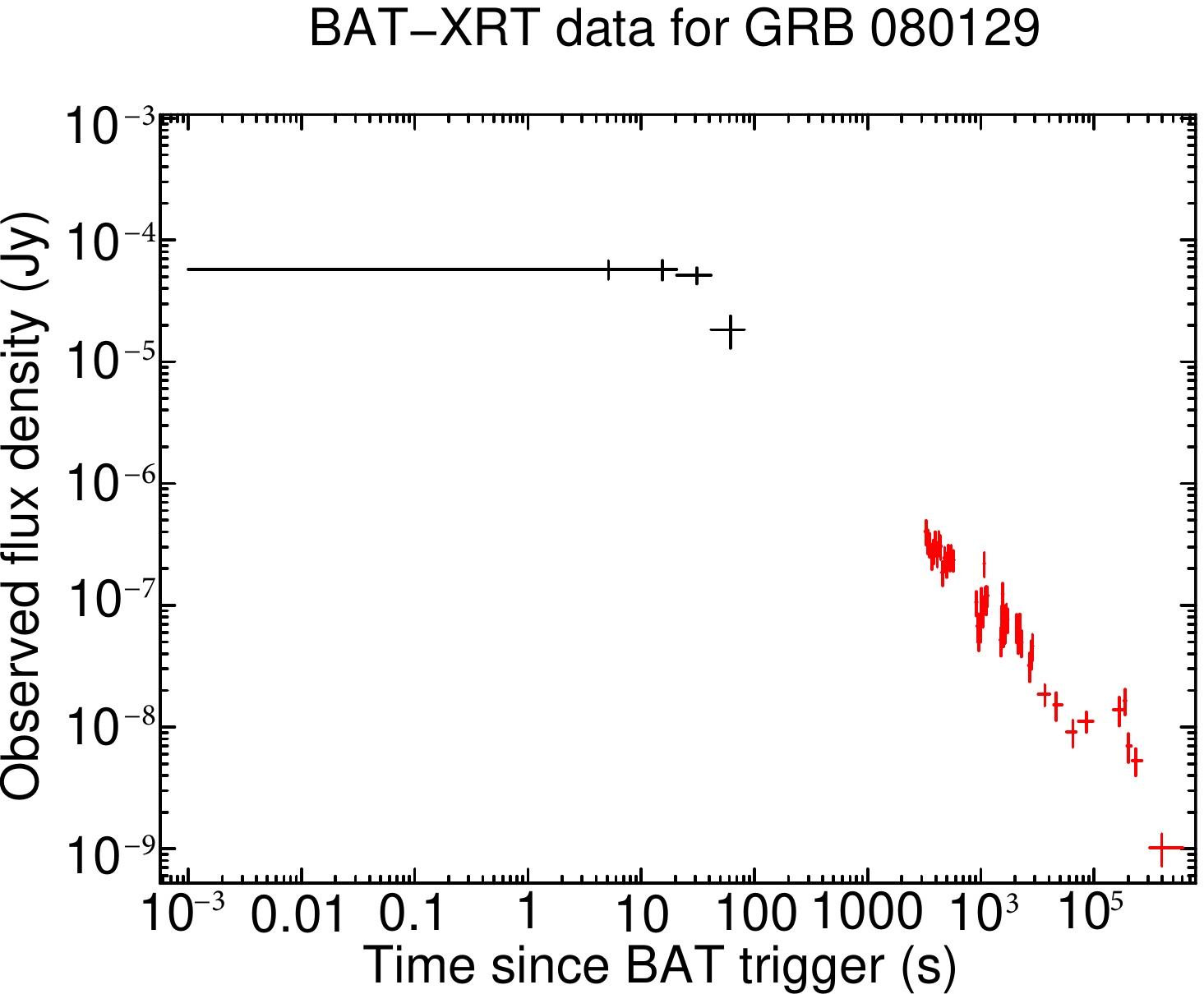}
\caption{\label{fig:grb080129}Swift observation of gamma and X-ray emission (lightcurve) from the GRB 080129. Black/red markers show gamma/X-ray flux. Source: UK Swift Science Data Centre}
\end{figure}

Swift's X-Ray Telescope (XRT) is designed to measure the GRB and afterglow fluxes, spectra, and lightcurves over a broad dynamic range, encompassing flux ranges greater than seven orders of magnitude. Within 10 seconds of target acquisition for a typical GRB, the XRT can pinpoint GRBs to within 5-arcsec. It can also study the X-ray counterparts of GRBs starting 20-70 seconds after the burst discovery and continuing for days to weeks \cite{xrt_2005,2000SPIE.4140...87H}. 
The telescope's energy resolution is around 260 eV and its energy range is $0.2 - 10$ keV, while it is often used for the $0.3 - 10.0$ keV region. We can see an example $\gamma$ and X-ray lightcurve plot on Fig. \ref{fig:grb080129}.

Swift BAT was triggered by more than 1600 GRBs up to the end of 2023 from which 1357 were detected by the XRT instrument simultaneously.
For our study, we used several observed and calculated data from both the BAT and XRT instruments: T90 duration,
fluence, 1sec peak flux from the gamma detectors and 11 hours flux, 24 hours flux, decay index, spectral index, and X HI column density from the X-ray instrument.

The BAT instrument gives critical data on the GRBs including the T$_{90}$ duration, which indicates the period when $90\%$ of the burst's total background-subtracted counts are identified. We also used the fluence, representing the total energy received in ergs per square centimetre and the 1-second peak photon flux, indicating the highest photon count rate in a single second during the burst.

In combination with the BAT data, the XRT database provides information about GRBs' afterglow phase. Key data points include 11- and 24-hour flux data, which follow the GRB's X-ray brightness after the first burst. We also used the initial temporal index and the spectral index,  
given by the XRT to characterise the afterglow decay rate and X-ray energy distribution, respectively. Furthermore, the intrinsic HI column density calculation provides an estimate of how much interstellar hydrogen the X-rays have gone through in the vicinity of the burst, revealing information about the environment around the GRB.

\subsection{Mathematical summary}

Canonical correlation assumes we have two sets of variables: $X$ and $Y$. The first set, $X$, contains $x_i=\{x_1,x_2,...,x_m\}$ and $Y$, the second one, $y_i=\{y_1, y_2, ...,y_r\}$ variables. We make n observations for each variable. Using the linear combination of the $X$ variables we develop

\begin{align}
U = \sum_{i=1}^m a_i\cdot x_i
\label{eq:U}
\end{align}

and using that of $Y$ 

\begin{align}
V = \sum_{i=1}^r b_i\cdot y_i
\label{eq:V}
\end{align}

variables and ask: how can be the $a$ and $b$ set of coefficients selected that the corr$(U,V)$, the correlation between $U$ and $V$ variables, obtained above, has the maximum value. For more details, see \cite{kendall1976advanced,Hardle2007-pk, Ignacio_Gonzalez_Sebastien_Dejean2007-vx}.

To maximise the correlation between $U$ and $V$, the coefficients $a= \{a_1, a_2, \ldots, a_k\} $ and $b = \{b_1, b_2, \ldots, b_l\} $ must meet the following criteria:

\begin{align}
\begin{bmatrix}
-\lambda \mathbf{R}_{xx} & \mathbf{R}_{xy} \\
\mathbf{R}_{xy} & -\lambda \mathbf{R}_{yy}
\end{bmatrix}
\begin{bmatrix}
\mathbf{a} \\
\mathbf{b}
\end{bmatrix}
= 0
\end{align}

$R_{xx}$ is the correlation matrix between the $x$ variables, $R_{yy}$ is the correlation matrix between the $y$ variables, and $R_{xy}$ is the x-y cross-correlation matrix. If the determinant is zero, then this equation has a non-trivial solution ($a$ and $b$ are not zero):

\begin{align}
\det \begin{bmatrix}
-\lambda \mathbf{R}_{xx} & \mathbf{R}_{xy} \\
\mathbf{R}_{xy} & -\lambda \mathbf{R}_{yy}
\end{bmatrix}
= 0
\end{align}

This determinant equation gives a polynomial equation for $\lambda^2$, with min$(k, l)$ solutions. The square root of each solution, $\lambda$, is known as the canonical correlation coefficient. Once $\lambda$ is known, the associated a and b coefficients can be calculated, defining a pair of canonical variables $U$ and $V$, also known as canonical factors.

We may readily comprehend the canonical factors by calculating the canonical factor structure matrix. We take the largest correlation and the corresponding $(U,V)$ pair and calculate their correlation with the input variables. The matrix produced in this manner contains information on the contribution of the input variables to the calculated canonical ones. We can continue similarly with the other canonical factors.
Based on the canonical factor structure, we can calculate the proportion of input variable variances explained by the computed canonical variables. The proportion of variance of variables defining $U$ explained by $V$, also known as redundancy of the $X$ set of variables with regard to $V$, and vice versa.

The generated canonical correlations should be checked for substantial departures from randomness. For this, we use Wilk's lambda parameter counting approach \cite{NATH1985297}. This procedure leads to a $\chi^2$ distribution test, which can be used to quickly determine the significance of the obtained result. The null hypothesis we employ is that the canonical variables $X$ and $Y$ are unconnected, or uncorrelated. The approach effectively applies a significance test to the collection of variables $X$ and $Y$.

\section{Results}

As written above, we examined 1357 GRBs that had BAT and XRT observations. From this dataset, however, we could only use those where all the parameters we are examining have been measured. Moreover, we examined the values of the 8 variables and about 200 data points seemed incorrect (with 3$\sigma$ confidence level). These values were too large or small (likely outlier). For this, we used the box-plot outlier search procedure. Finally, we have 930 points of observation left. It should be noted that the logarithm of most parameters (duration, fluence, peak flux, N(HI), X-ray flux) should be used in the procedure, as this is the only way to ensure that the correlation is derived from the bulk of the data, as it brings the data distribution closer to normal.

After the data of the variables were cleaned, the remaining variables were separated into two sets and we have determined the number of dimensions (canonical variables) that are significant in explaining the association between the 2 sets of variables ($\gamma$ and X-ray data).

\begin{figure*}
\includegraphics[width=0.8\linewidth]{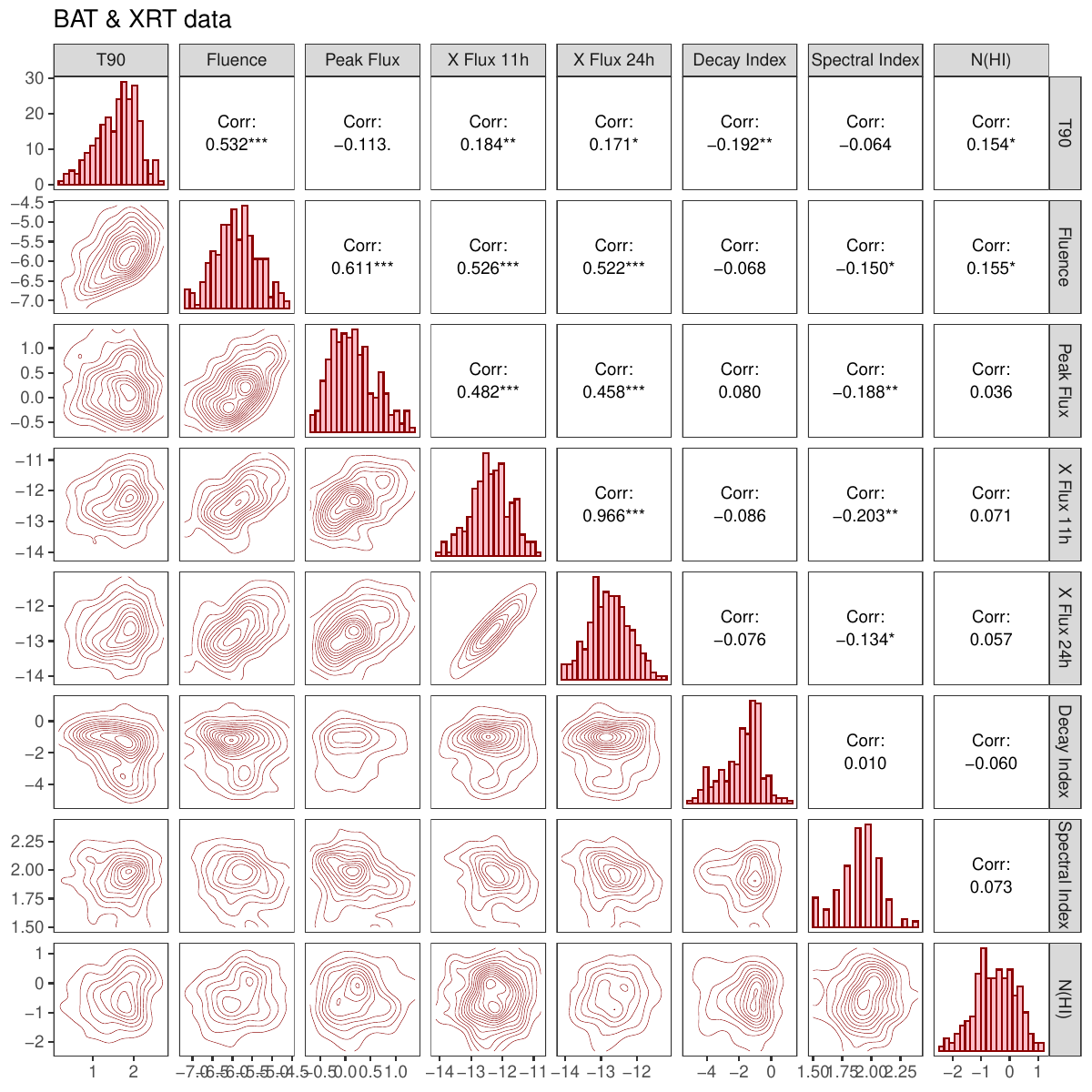}
\caption{Cross-correlation matrix of the GRBs' parameters ('outliers' excluded). The number of asterisks indicates the significance of the correlation.}
\label{fig:cross_corr}
\end{figure*}

Then, we were able compute the correlation matrix within and between the two datasets and visualised the variables in X and Y variables to see how they relate to each other. These cross-correlation plots are shown in Fig.~\ref{fig:cross_corr}. The figure shows the correlations between the different parameters, as well as the significance of these correlations with asterisks. The X-ray fluxes show the strongest correlation between each other, which is completely obvious. However, a weak but highly significant correlation can also be seen between the X-ray 11h flux and the gamma T90 parameters. A similar correlation with duration can be seen with the intrinsic hydrogen column density. Another observable fact is that there is a moderately strong correlation between gamma fluence and X-ray flux. This pairwise plots clearly showed that for both X and Y matrices, there is some visible correlation between variables. In this situation, canonical correlation analysis is ideal for exploring the overall relationship between the two sets.

In the next step, we calculated the canonical correlation coefficients for the two groups as above. The variable U was calculated from the BAT data, while the variable V was derived from the X-ray data. Next, we examined the relationship between these canonical variables and compared them with previous results.

In our analysis, we evaluated the relationships between canonical variables to examine how gamma-ray bursts' (GRBs) gamma-ray and X-ray properties interact. The canonical correlation analysis revealed strong associations, with Wilks' Lambda values providing insight into the significance of these associations.

From our Wilks' Lambda analysis, for the full dataset, the test for the first three canonical correlations gave a Lambda of $0.527$ ($F = 44.22$, $p<10^{-10}$) from the first to third canonical correlation, indicating a strong association between the GRB gamma and X-ray variables across multiple dimensions. The second-to-third canonical test yielded a Lambda of $0.960$ ($F = 4.75$, $p<0.001$), showing that even as we move to higher canonical dimensions, a notable relationship persists between the variable sets, albeit with a slightly reduced significance. Finally, the third canonical test gave a Lambda of $0.992$ ($F = 2.57$, $p=0.053$), suggesting a weak, yet non-trivial link between the variable sets in these later dimensions.

\section{Discussion}

The investigation of the canonical variables produced from the canonical correlation analysis sheds light on the link between gamma-ray bursts' optical and X-ray features. Tables~\ref{tab:cancor_old} and~\ref{tab:cancor} highlight the important findings from the BAT and XRT datasets. The correlation coefficients between each pair of canonical variables $U$ and $V$ help us to understand the relationships between the two sets of variables.
It is vital, to again, notice that the canonical variables U are the gamma, whereas the V coefficients are derived from the X-ray data.

Our results support the conclusion that while the first canonical dimension (U1) is strongly associated with the main GRB properties, such as $T90$, fluence, and peak flux, suggesting a primary linkage between these gamma-ray properties and the X-ray afterglow’s evolution, subsequent canonical variables (U2, U3) reveal increasingly subtle relationships, capturing more nuanced aspects of the GRB dynamics and potentially different phases of the bursts. This graded strength of association through successive canonical dimensions reflects the complexity of the GRB processes, and the Lambda values underscore the decreasing influence of higher canonical variables on the explaining variance in the data.

\begin{table*}
\begin{tabular}{lrrr|}
 & T90 & Fluence & PeakFlux\\ \hline
U1 & \textbf{-0.520} & \textbf{-0.911} & \textbf{-0.785}\\
U2 & \textbf{0.853} & {0.094} & \textbf{-0.605}\\
U3 & {0.065} & \textbf{0.430} & -0.148
\end{tabular}
\begin{tabular}{|lrrrrr}
 & 11hFlux & 24hFlux & Decay ind. & Sp. index & N(HI)\\ \hline
V1 & \textbf{-0.927} & \textbf{-0.825} & {0.091} & \textbf{0.273} & \textbf{-0.206}\\
V2 & -0.012 & -0.020 & \textbf{-0.856} & \textbf{0.257} & \textbf{-0.424}\\
V3 & {-0.109} & {-0.066} & \textbf{-0.409} & \textbf{-0.706} & 
\textbf{0.228}
\end{tabular}
\caption{The connection between the U (from BAT data) and V (from XRT data) canonical correlation coefficients and the GRB parameters using only the data before the July of 2008 for the comparison. Bold-face indicates the 3$\sigma$ significance level.}
\label{tab:cancor_old}
\end{table*}

\begin{table*}
\begin{tabular}{lrrr|}
 & T90 & Fluence & PeakFlux\\ \hline
U1 & \textbf{-0.611} & \textbf{-0.946} & \textbf{-0.763}\\
U2 & \textbf{0.788} & \textbf{0.120} & \textbf{-0.642}\\
U3 & {-0.092} & \textbf{0.324} & -0.088
\end{tabular}
\begin{tabular}{|lrrrrr}
 & 11hFlux & 24hFlux & Decay ind. & Sp. index & N(HI)\\ \hline
V1 & \textbf{-0.942} & \textbf{-0.898} & \textbf{0.131} & \textbf{0.216} & \textbf{-0.216}\\
V2 & -0.083 & -0.028 & \textbf{-0.916} & {-0.023} & \textbf{-0.439}\\
V3 & \textbf{-0.316} & \textbf{-0.522} & -0.085 & \textbf{-0.708} & 
\textbf{0.121}
\end{tabular}
\caption{The connection between the U (from BAT data) and V (from XRT data) canonical correlation coefficients and the GRB parameters, based on the entire dataset up to the end of 2023. Boldface indicates the 3$\sigma$ significance level.}
\label{tab:cancor}
\end{table*}

According to the Table~\ref{tab:cancor}, the first canonical variable $(U1)$ has a substantial negative correlation with $T_{90}$ ($r = -0.611$), fluence ($r = -0.946$), and peak flux ($r = -0.763$). This suggests that the parameters determined from the BAT data are strongly connected to the first canonical axis, implying a link between the length and intensity of gamma-ray bursts and the early stages of the X-ray afterglow. The second canonical variable ($U2$) has a substantial connection with $T_{90}$ $(r = 0.788)$, but less so with fluence $(r = 0.120)$, and in the opposite direction with peak flux $(r = 0.642)$, showing a more complicated dynamic between different elements of the gamma-ray bursts.

For the $V$ variables, the first canonical variable $(V1)$ has a large negative correlation with the 11-hour flux $(r = -0.942)$ and the 24-hour flux $(r = -0.898)$, whereas the spectral index and HI column density have lesser, but still significant relationships. This implies that the early X-ray fluxes and their spectrum features acquired from the XRT data are strongly connected to the optical properties of the gamma-ray bursts, supporting the close relationship between the BAT and XRT data. Further investigation of these findings can aid in understanding the links between the distinct phases of gamma-ray bursts and the processes of the central engine.

The Table~\ref{tab:cancor} also shows other examples of weak, but extremely significant correlations, which are highlighted in bold. These substantial correlations, despite their modest levels, indicate that even minor interactions between variables can be statistically significant. For example, the spectral index and HI column density have modest, but substantial correlations with the first canonical variable $V1$, showing that these parameters, while not highly associated, play an important role in the overall relationship between the BAT and XRT datasets. These findings emphasise the necessity of taking into account even slight correlations when evaluating the intricate interplay between the distinct phases of gamma-ray bursts.

Furthermore, the second and third canonical variates ($U2$, $U3$, and $V2$, $V3$) reveal more intricate interactions, shedding information on how many components of the GRB development interact. The substantial association of $U2$ with $T_{90}$ and peak flux, but not with fluence, and $V2$ with the decay index imply that these variables represent the developing dynamics of the GRB as it transitions from the prompt emission to the afterglow phase. These findings are critical for creating a complete model of the GRB behaviour and deepening our knowledge of the fundamental mechanisms that drive such catastrophic occurrences.

The term 'complete model' here refers to a model that incorporates the many stages of the GRB event as well as the physical characteristics involved, including the connection between the prompt emission and the afterglow that follows.
Such a model aims to explain the physical processes that take place during the explosion, including the dynamics and interactions between the different 
components of the observation (e.g. T90 duration, peak flux, and decay index). This model can be particularly useful for groups of GRBs with significant variations in spectral energy distribution between the early and late phases, i.e. not limited to a specific type, but rather covering the entire GRB phenomena.

Overall, the canonical correlation analysis provides a strong framework for investigating the complex correlations between the GRB observational features. The substantial connections found between the canonical variables highlight the interrelated nature of GRBs' prompt and afterglow phases, providing crucial information for future theoretical and observational studies. Future research might improve on these findings by including more factors and broadening the dataset, refining our understanding of the fundamental processes that underlie GRB events.

\subsection{Comparison with previous results}

Previous comparable investigations such as the one presented at the GAMMA-RAY BURST: Sixth Huntsville Symposium in 2008 have shed light on the relationships between gamma-ray bursts and their observational features. The 2008 research produced many important findings:

\begin{enumerate}[label=(\alph*)]
\item \,In comparison to the X-ray decay index and spectral index, the gamma-ray fluence and early X-ray flux were shown to be the most significant contributions to the canonical correlation.
\item \,The HI column density contributed significantly to the canonical correlation. This discovery was interpreted in terms of the collapsar hypothesis for long GRBs, implying that the progenitor's ejection of an HI envelope was critical in creating the GRB.
\end{enumerate}

Our recent investigation confirms these previous findings, emphasising the importance of the gamma-ray fluence and early X-ray flux in the canonical correlation analysis. The gamma-ray fluence and early X-ray flux continue to make significant contributions to the canonical correlations, as seen by the boldface coefficients in Table \ref{tab:cancor}. These findings are consistent with prior conclusions, emphasising the significance of these characteristics in understanding the link between GRBs' prompt and afterglow phases.

Our investigation found that the HI column density made a significant contribution to the canonical correlation, as seen in the last row of Table \ref{tab:cancor}. This conclusion is consistent with the 2008 study and lends credence to the collapsar concept of long GRBs. The strong association between the HI column density and canonical factors shows that the GRB progenitor's ejection of an HI envelope plays an important role in the observed X-ray afterglow features. This consistent finding across investigations emphasises the reliability of the HI column density as a critical parameter in a GRB study.

Furthermore, our approach has offered a better understanding of the canonical relationships with other factors. For example, while the spectral index and X-ray decay index make smaller contributions, they nonetheless play a role in the overall correlation structure. This detailed knowledge helps to enhance our models of GRB behaviour and promotes additional research into the interactions of many observable features.

\begin{figure*}
\includegraphics[width=0.49\linewidth]{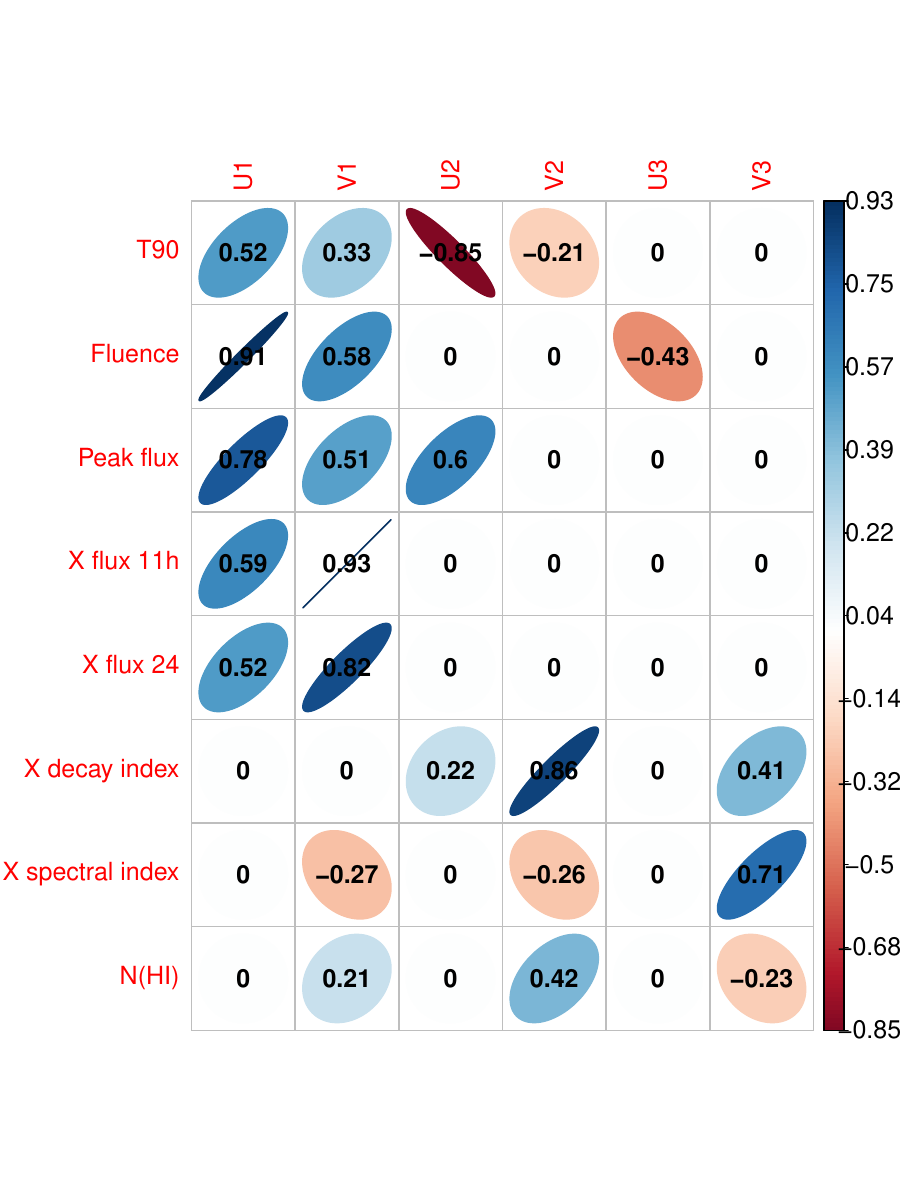}
\includegraphics[width=0.49\linewidth]{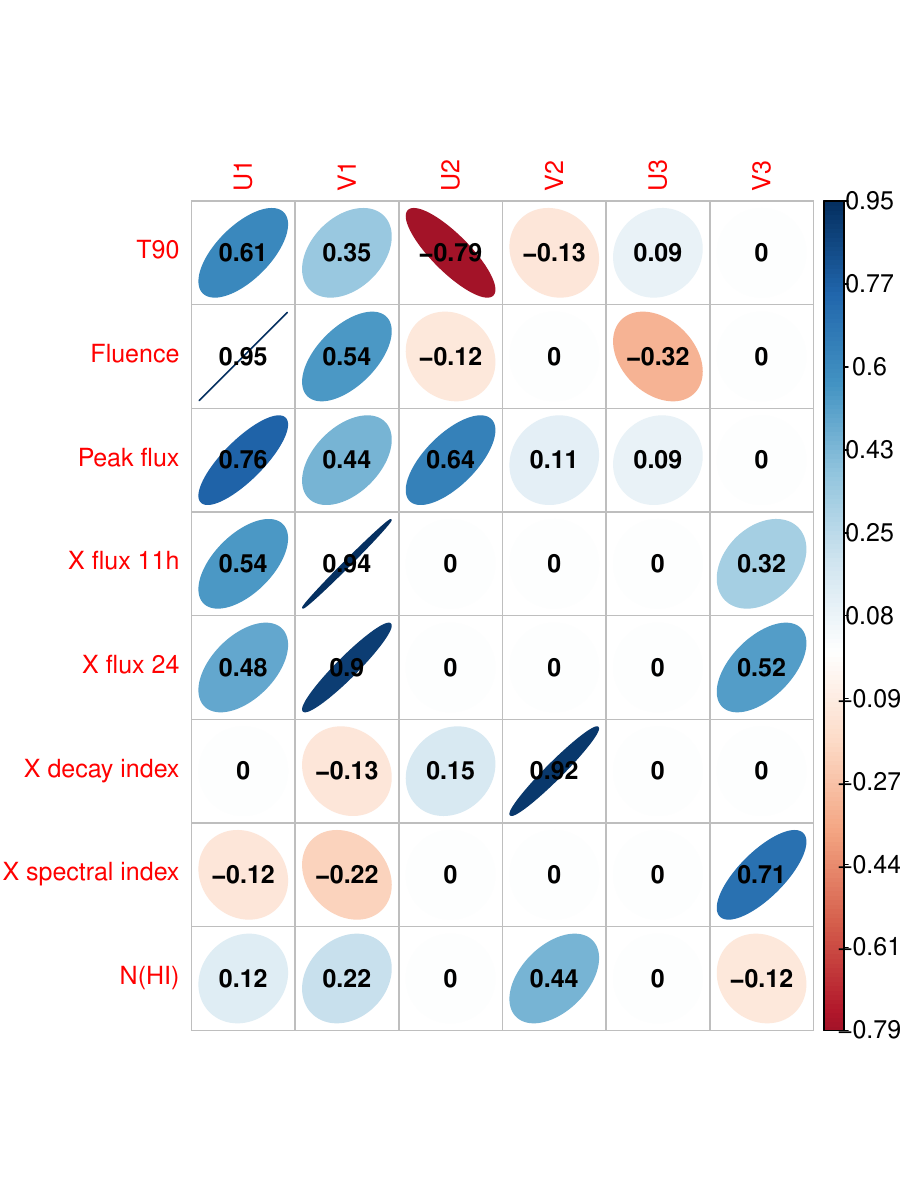}
\caption{The correlations between the canonical variables U and V and the GRB parameters in the 2008 research (left) and the current investigation up to 2023 (right). The graphs provide the correlations and significance levels for each parameter, allowing for a visual comparison of the results. The colour blue denotes a positive association, whereas red shows a negative correlation. The brightness of the colour and the flatness of the circle indicate the strength of the association. The image displays only relationships with a significance level of 3$\sigma$.}
\label{fig:cca_comp}
\end{figure*}

Figure \ref{fig:cca_comp} depicts the correlations between the canonical variables U and V and the GRB parameters for both the 2008 study and our current investigation up to 2023. The plots in the image indicate the correlations and significance levels of each parameter, allowing for a visual comparison of the results. The consistent patterns found in the two datasets support the canonical correlation analysis and the significance of the identified factors in the GRB study.

Overall, the comparison of earlier data reveals a remarkable consistency in the major discoveries regarding GRB characteristics. Here, of course, we note that the strength of the correlation between the variables $N(HI)$ and $V2$ has lost significance. Alternatively, the now computed $V3$ has changed sign, which in the case of the coefficient shows the opposite direction. Moreover, this $V3$ shows even stronger correlations now than in 2008. To summarise, Gamma-ray fluence, early X-ray flux, and HI column density have all been shown to provide significant contributions to canonical correlations, emphasising their relevance in GRB research. These findings not only support previous studies, but also pave the way for further investigations into the intricate systems behind GRBs.

\subsection{Changing X-ray spectra parameters}

In this work, we investigated the X-ray photons of the GRB. The correlation between the parameters obtained from the X-ray spectra and the canonical variables is well established. In the next work, we will try to refine the spectrum obtained from X-ray photons.

\begin{figure*}
\begin{center}
\includegraphics[width=0.505\linewidth, trim= 0cm -0.6cm 0cm 0cm]{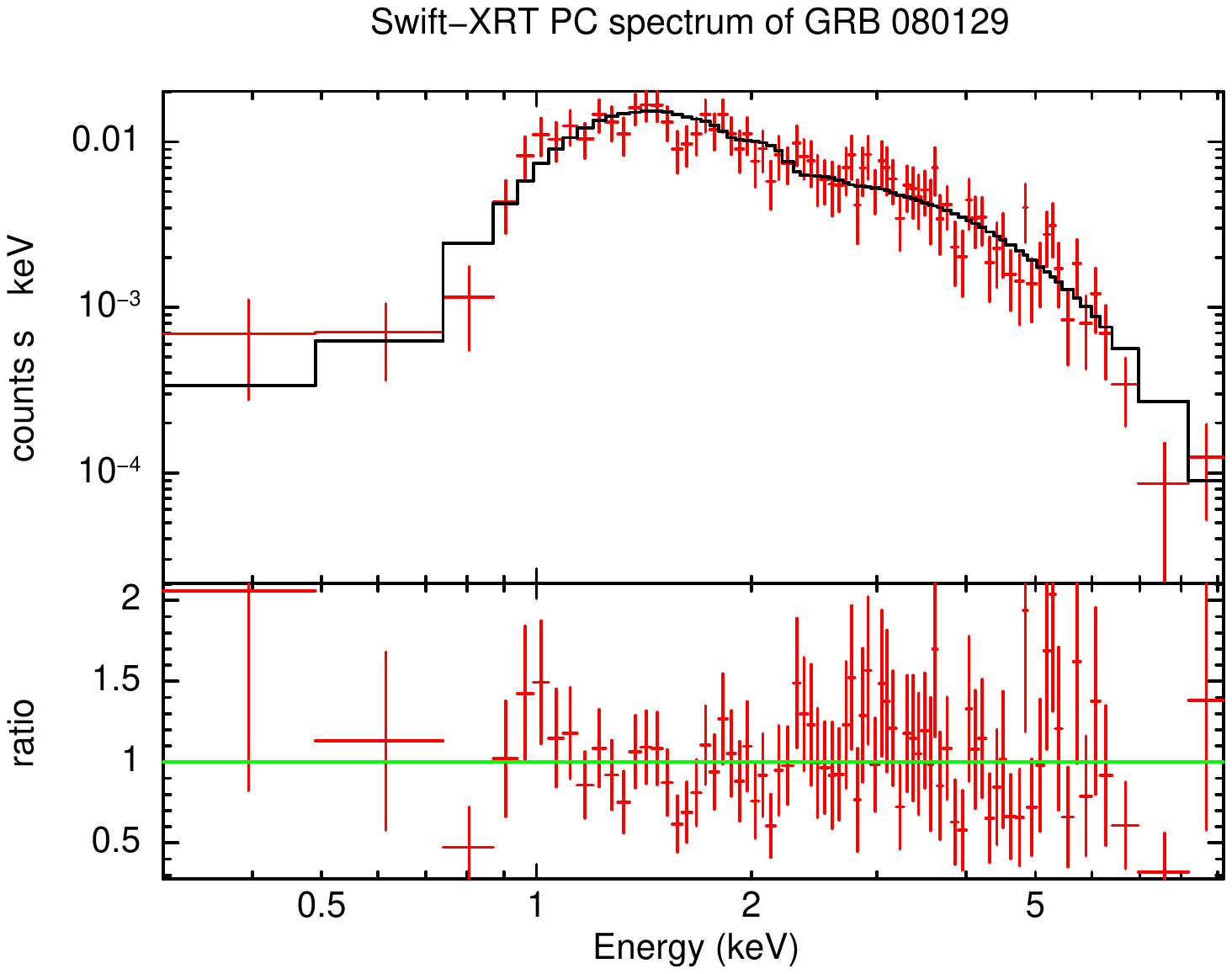} \includegraphics[width=0.443\linewidth, trim= 2cm 0 0 0]{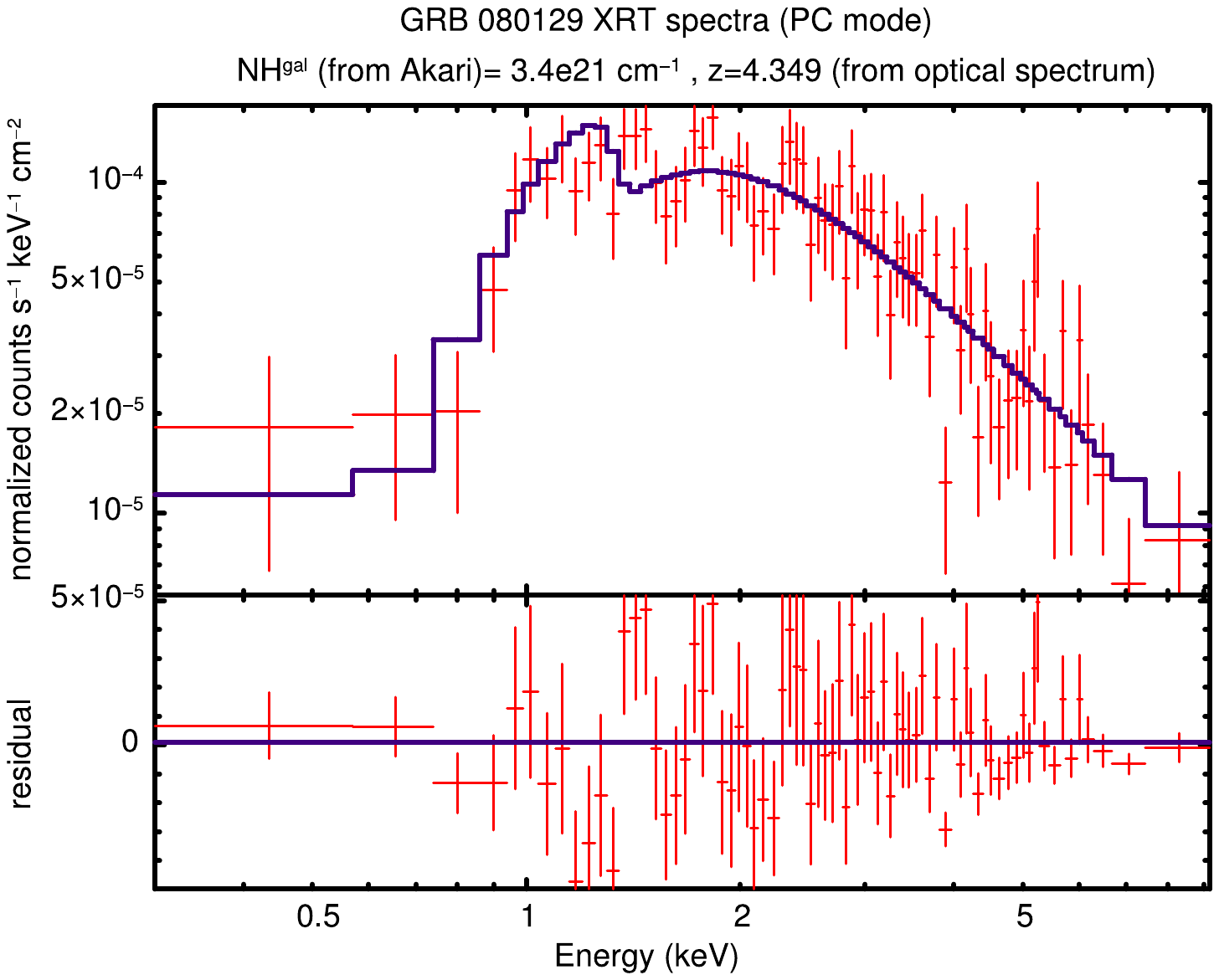}
\caption{\label{fig:grb080129_spetrum}Left: Time-averaged X-ray spectra (UK Swift Science Data Centre (UKSSDC), Evans et al., 2009 \cite{evans2009}) Right: Our previous result from Racz et al., 2017 \cite{racz_2017coska}, the GRB080129's refitted XRT spectrum with the new values (foreground N(H) and redshift). The significant difference between the two fits is also easily visible to the eye.}
\end{center}
\end{figure*}

Our previous results show that in many cases, the internal hydrogen column density can change significantly during the spectrum fitting procedure \cite{racz_2017coska,2019MNRAS.489.3778D}. A very good example of it can be seen in Fig.~\ref{fig:grb080129_spetrum}.

We plan to carry out the following steps to use higher quality input data for the spectral fitting:

\begin{enumerate}[label=(\alph*)]
\item \,Unified settings and models on all GRB's spectra
\item \,Only (reliable) spectroscopic redshifts
\item \,Use more/newer redshift data 
\item \,Higher resolution Galactic foreground HI density maps
 calculated from Planck maps
\item \,Initial values are important for a correct iteration
\end{enumerate}

To date, the Swift has observed nearly 300 GRBs for which reliable X-ray spectra have been obtained by the Swift XRT instrument, and they have spectroscopically measured redshifts. These redshifts are often not very accurate at first, so significant differences may be found compared to the initial data after a few days. Additionally, subsequently released data do not always get incorporated into the automatic processing procedures. 

For all redshift measurements described in b and c, we need to check the value in the published databases and then verify their validity in the GCN notes. In addition, we are also planning to use a reliable value not found in the databases, but published in the literature. The precomputation mentioned in 'd' will have to be done as follows. As it was already shown from ISO measurements, the FIR based estimation of Galactic ISM column densities must include measurements at wavelengths beyond $100~ \mu $m (see eg. \cite{Toth2000}, \cite{Toth2004}). Accordingly, we used the 5 arcmin resolution Planck visible V-band renormalised (Av\_RQ) extinction map\footnote{COM\_CompMap\_Dust-DL07-AvMaps\_2048\_R2.00.fits} as a starting point estimating the Galactic foreground column density, that is based on the multiband Planck sky survey corrected using visual extinctions of quasars from the Sloan Digital Sky Survey \cite{2016A&A...586A.132P}. From this whole-sky extinction field image, we extracted the dust extinction for each GRB. We note that an alternative estimate could be based on NIR extinction of stars \cite{Cambresy2013}. Assuming that the dust is properly mixed in the interstellar material \cite{2019PASJ...71...10T}, we can calculate the Galactic hydrogen column density using \cite{2009MNRAS.400.2050G}:
\begin{align}
N(H)=2.21 (\pm 0.09 )\cdot 10^{21} \times A_v
\label{eq:nh}
\end{align}
where $A_v$ is the visual extinction read from Planck images. We note here that there is a variation of extinction to gas density in the Galactic ISM (see eg. \cite{Toth1995}, \cite{Harjunpaa2005}. The resulting hydrogen column density values can be used directly in the re-fitting of the spectra.

\begin{figure}
    \centering
    \includegraphics[width=1\linewidth]{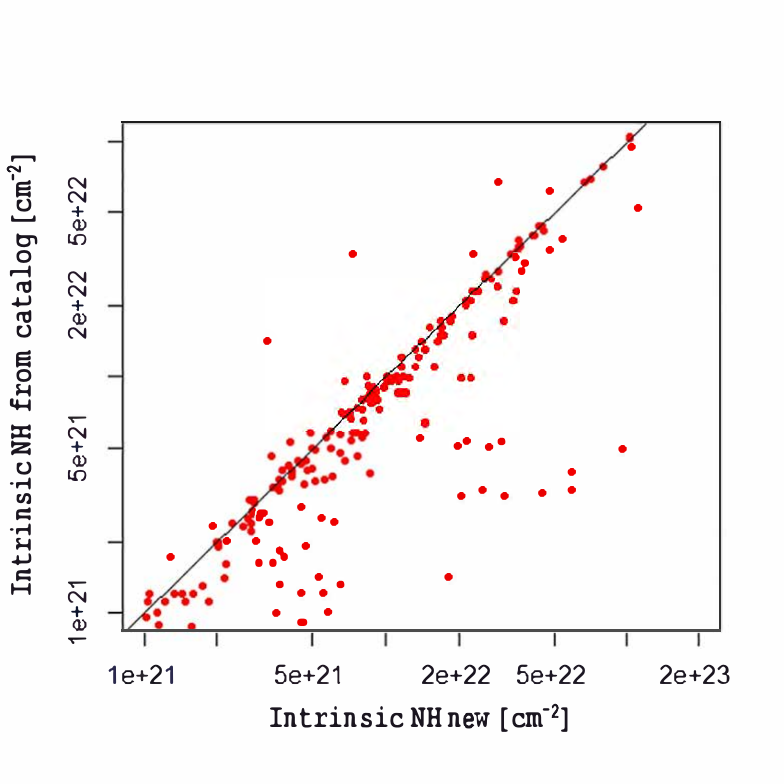}   
    \caption{\label{fig:nh_change}The variation of the hydrogen column density for selected high N(H) GRBs in $\mathrm{cm}^{-2}$. The vertical axis shows the N(H) value from the catalogue, while the horizontal axis shows our calculated column densities.}
\end{figure}

For some well-detected GRB cases with a high hydrogen column density, we performed refitting of the spectra. By re-fitting the X-ray spectra of the GRBs, it is clear how divergent the intrinsic hydrogen column density is (Fig.~\ref{fig:nh_change}). In dozens of cases, we got discrepancies of orders of magnitude. On average, we got a value $4.5\cdot10^{21} cm^{-2}$ higher than the value in the official catalog, which corresponds to an increase of approximately $50\%$.

With the new data, we will have to carry out the entire investigation again in order to check the robustness of the obtained correlations and to find the previously hidden connections. But this work is beyond the scope of this article.

\section{Conclusions}

We used canonical correlation analysis to investigate the relationship between gamma (fluence, 1 sec peak flux, and duration) and X-ray data (11 hours flux, 24 hours flux, decay index, spectral index, and HI column density). Using the canonical variables obtained from the study, we calculated their correlations (canonical loadings) with the originals. The canonical loadings indicated that the gamma-ray fluence and early X-ray flux contribute the most to the correlation, as opposed to the X-ray decay index and spectral index. 

An interesting finding appears to be that the HI column density contributes significantly to the connection. Accepting the collapsar model of long GRBs, this phenomenon may suggest that the stellar wind of the massive progenitor star prior to the burst had created a substantial HI envelope that was already present at the moment of the GRB. We compared the results to previous studies and found comparable results; the previous associations remained significant in the same way. 

Because, as previously demonstrated, the precision of fitting to the X-ray spectra has a significant impact on the calculation of the inherent column density. Thus, in this article and in a few circumstances, we have investigated this phenomena and proposed a solution. However, this study is still continuing and requires further investigation.

\begin{acknowledgements}
The authors would like to thank the Hungarian TKP2021-NVA-16 programme for their support. Supported by the EKÖP-24-4-II-48 University Research Scholarship Program of the Ministry for Culture and Innovation from the source of the National Research, Development and Innovation Fund. The authors would like to thank L. Viktor Tóth and Zsolt Bagoly for their help in the preparation of this article.
\end{acknowledgements}

\bibliographystyle{actapoly}
\bibliography{bib}

\end{document}